\documentclass[aps,prl,showpacs,groupedaddress,superscriptaddress,twocolumn]{revtex4}
\usepackage{amsfonts}
\usepackage{graphicx}
\usepackage{latexsym}
\usepackage{makeidx}
\usepackage{amsmath}
\usepackage{amssymb}
\usepackage{graphicx}

\begin{document}

\title{Robustness of quantum discord to sudden death}
\author{T. Werlang}
\affiliation{Departamento de F\'{\i}sica, Universidade Federal de S\~{a}o Carlos, P.O.
Box 676, CEP 13565-905, S\~{a}o Carlos\textit{, }S\~{a}o Paulo\textit{, }%
Brazil}
\author{S. Souza}
\affiliation{Departamento de F\'{\i}sica, Universidade Federal de S\~{a}o Carlos, P.O.
Box 676, CEP 13565-905, S\~{a}o Carlos\textit{, }S\~{a}o Paulo\textit{, }%
Brazil}
\author{F. F. Fanchini}
\affiliation{Instituto de F\'{\i}sica Gleb Wataghin, Universidade Estadual de Campinas,
P.O. Box 6165, CEP 13083-970, Campinas, S\~{a}o Paulo, Brazil}
\author{C. J. Villas-B\^{o}as}
\email{celsovb@df.ufscar.br}
\affiliation{Departamento de F\'{\i}sica, Universidade Federal de S\~{a}o Carlos, P.O.
Box 676, CEP 13565-905, S\~{a}o Carlos\textit{, }S\~{a}o Paulo\textit{, }%
Brazil}
\affiliation{Max-Planck-Institut f\"{u}r Quantenoptik, Hans-Kopfermann-Str. 1, D-85748,
Garching, Germany}

\begin{abstract}
We calculate the dissipative dynamics of two-qubit quantum discord under
Markovian environments. We analyze various dissipative channels such as
dephasing, depolarizing, and generalized amplitude damping, assuming
independent perturbation, in which each qubit is coupled to its own channel.
Choosing initial conditions that manifest the so-called sudden death of
entanglement, we compare the dynamics of entanglement with that of quantum
discord. We show that in all cases where entanglement suddenly disappears,
quantum discord vanishes only in the asymptotic limit, behaving similarly to
individual decoherence of the qubits, even at finite temperatures. Hence,
quantum discord is more robust than the entanglement against to decoherence
so that quantum algorithms based only on quantum discord correlations may be
more robust than those based on entanglement.
\end{abstract}

\maketitle

Entanglement is widely seen as the main reason for the computational
advantage of quantum over classical algorithms. This view is backed up by
the discovery that, in order to offer any speedup over a classical computer,
the universal pure-state quantum computer would have to generate a large
amount of entanglement \cite{vidal}. However, quantum entanglement is not
necessary for deterministic quantum computation with one pure qubit (DQC1),
introduced by Knill and Laflamme \cite{knill}. As seen in \cite{datta,white}%
, although there is no entanglement, other kinds of nonclassical correlation
are responsible for the quantum computational efficiency of DQC1. Such
correlations are characterized as quantum discord \cite{zurek}, which
accounts for all nonclassical correlations present in a system, being the
entanglement a particular case of it. Besides its application in DQC1,
quantum discord has also been used in studies of quantum phase transition 
\cite{phase transition}, estimation of quantum correlations in the Grover
search algorithm \cite{Cui} and to define the class of initial system-bath
states for which the quantum dynamics is equivalent to a completely positive
map \cite{lidar}.

When considering a pair of entangled qubits exposed to local noisy
environments, disentanglement can occur in a finite time \cite{diosi03,
citeESD, eberly, yu, davidovich}, differently from the usual local
decoherence in asymptotic time. The occurrence of this phenomenon, named
\textquotedblleft entanglement sudden death\textquotedblright\ (ESD),
depends on the system-environment interaction and on the initial state of
the two qubits. Our goal was to investigate the dynamics of quantum discord
of two qubits under the same conditions in which ESD can occur. We show in
this letter that even in cases where entanglement suddenly disappears,
quantum discord decays only in asymptotic time. Furthermore, this occurs
even at finite temperatures. In this sense, quantum discord is more robust
against decoherence than entanglement, implying that quantum algorithms
based only on quantum discord correlations are more robust than those based
on entanglement.

There are various methods to quantify the entanglement between two qubits 
\cite{bennett96, wootters98, vidal02} and, even when they give different
results for the degree of entanglement of a specific state, all of them
result in $0$ for separable states. Therefore, under dissipative dynamics
where an initial entangled state can disappear suddenly, all of them
necessarily agree about the time when the quantum state becomes separable.

Here, to investigate the two-qubit entanglement dynamics we use concurrence
as the quantifier \cite{wootters98}. The concurrence is given by $\max
\left\{ 0,\Lambda \left( t\right) \right\} $, where $\Lambda (t)={\lambda
_{1}}-{\lambda _{2}}-{\lambda _{3}}-{\lambda _{4}}$ and $\lambda _{1}\geq
\lambda _{2}\geq \lambda _{3}\geq \lambda _{4}$ are the square roots of the
eigenvalues of the matrix $\rho (t)\sigma _{2}\otimes \sigma _{2}\rho ^{\ast
}(t)\sigma _{2}\otimes \sigma _{2}$, $\rho ^{\ast }(t)$ being the complex
conjugate of $\rho (t)$ and $\sigma _{2}$ the second Pauli matrix. The
density matrix we use to evaluate concurrence has an $\mathrm{X}$ structure 
\cite{yu}, defined by $\rho _{12}=\rho _{13}=\rho _{24}=\rho _{34}=0$, which
are constant during the evolution, for the various dissipative channels used
here. In this case, the concurrence has a simple analytic expression $%
C\left( t\right) =2\max \left\{ 0,\Lambda _{1}\left( t\right) ,\Lambda
_{2}\left( t\right) \right\} \,$, where $\Lambda _{1}\left( t\right)
=\left\vert \rho _{14}\right\vert -\sqrt{\rho _{22}\rho _{33}}$ and $\Lambda
_{2}\left( t\right) =\left\vert \rho _{23}\right\vert -\sqrt{\rho _{11}\rho
_{44}}$. However, as pointed out above, entanglement is not the only kind of
quantum correlation. In quantum information theory, the Von Neumann entropy, 
$S\left( \rho \right) =-\mathrm{Tr}\left( \rho \log \rho \right) $, is used
to quantify the information in a generic quantum state $\rho $. The total
correlation between two subsystems $\mathcal{A}$ and $\mathcal{B}$ of a
bipartite quantum system $\rho _{\mathcal{A}\mathcal{B}}$ is given by the
mutual information, 
\begin{equation}
\mathcal{I}(\rho _{\mathcal{A}\mathcal{B}})=S(\rho _{\mathcal{A}})+S(\rho _{%
\mathcal{B}})-S(\rho _{\mathcal{A}\mathcal{B}}),  \label{Tcorrelation}
\end{equation}%
where $S(\rho _{\mathcal{A}\mathcal{B}})=-\mathrm{Tr}\left( \rho _{\mathcal{A%
}\mathcal{B}}\log \rho _{\mathcal{A}\mathcal{B}}\right) $ is the joint
entropy of the system \cite{nielsen}. This bipartite quantum state $\rho _{%
\mathcal{A}\mathcal{B}}$ is a hybrid object with both classical and quantum
characteristics and, in order to reveal the classical aspect of correlation,
Henderson and Vedral suggested that correlation could also be split into two
parts, the quantum and the classical \cite{vedral}. The classical part was
defined as the maximum information about one subsystem that can be obtained
by performing a measurement on the other subsystem. If we choose a complete
set of projectors $\left\{ \Pi _{k}\right\} $ to measure one of the
subsystems, say $\mathcal{B}$, the information obtained about $\mathcal{A}$
after the measurement resulting in outcome $k$ with probability $p_{k}$, is
the difference between the initial and the conditional entropy \cite{vedral} 
\begin{equation}
\mathcal{Q}_{\mathcal{A}}(\rho _{\mathcal{A}\mathcal{B}})=\max_{\left\{ \Pi
_{k}\right\} }[S(\rho _{\mathcal{A}})-\sum_{k}p_{k}S(\rho _{\mathcal{A}|k})],
\label{cc}
\end{equation}%
where $\rho _{\mathcal{A}|k}=\mathrm{Tr}_{\mathcal{B}}(\Pi _{k}\rho _{%
\mathcal{A}\mathcal{B}}\Pi _{k})/\mathrm{Tr}_{\mathcal{AB}}(\Pi _{k}\rho _{%
\mathcal{A}\mathcal{B}}\Pi _{k})$ is the reduced state of $\mathcal{A}$
after obtaining the outcome $k$ in $\mathcal{B}$. This measurement of
classical correlation assumes equal values, irrespective of whether the
measurement is performed on the subsystem $\mathcal{A}$ or $\mathcal{B}$,
for all states $\rho _{\mathcal{A}\mathcal{B}}$ such that $S(\rho _{\mathcal{%
A}})=S(\rho _{\mathcal{B}})$ \cite{vedral}. This condition is true of all
density operators used in this paper since they can be written as $\rho =%
\frac{1}{4}\left[ \mathbb{I}+c_{0}\left( \sigma _{3}\otimes \mathbb{I}+%
\mathbb{I}\otimes \sigma _{3}\right) +\sum_{j}c_{j}\sigma _{j}\otimes \sigma
_{j}\right] $, where $c_{i}$ and $\sigma _{i}$ $(i=1,2,3)$ are real
constants and Pauli matrices, respectively. Therefore $\rho _{\mathcal{A}%
}=\rho _{\mathcal{B}}$.

In this scenario, a quantity that provides information on the quantum
component of the correlation between two systems can be introduced as the
difference between the total correlations in (\ref{Tcorrelation}) and the
classical correlation in (\ref{cc}). This quantity is identical to the
definition of quantum discord introduced by Ollivier and Zurek in \cite%
{zurek}, namely%
\begin{equation}
D\left( \rho _{\mathcal{A}\mathcal{B}}\right) ={\mathcal{I}}\left( \rho _{%
\mathcal{A}\mathcal{B}}\right) -\mathcal{Q}\left( \rho _{\mathcal{A}\mathcal{%
B}}\right) ,  \label{D}
\end{equation}%
this being zero for states with only classical correlations \cite%
{zurek,vedral} and nonzero for states with quantum correlations. Moreover,
quantum discord includes quantum correlations that can be present in states
that are not entangled \cite{zurek}, revealing that all the entanglement
measurements such as concurrence, entanglement of formation, etc, do not
capture the whole of quantum correlation between two mixed separate systems.
For pure states, the discord reduces exactly to a measure of entanglement,
namely the entropy of entanglement.

In order to calculate the quantum discord between two qubits subject to
dissipative processes, we use the following approach. The dynamics of two
qubits interacting independently with individual environments is described
by the solutions of the appropriate Born-Markov-Lindblad equations \cite%
{carmichael}, that can be obtained conveniently by the Kraus operator
approach \cite{kraus,nielsen}. Given an initial state for two qubits $\rho
(0)$, its evolution can be written compactly as 
\begin{equation}
\rho (t)=\Sigma _{\mu ,\nu }E_{\mu ,\nu }\rho (0)E_{\mu ,\nu }^{\dagger },
\label{rho}
\end{equation}%
where the so-called Kraus operators $E_{\mu ,\nu }=E_{\mu }\otimes E_{\nu }$ 
\cite{nielsen} satisfy $\Sigma _{\mu ,\nu }E_{\mu ,\nu }^{\dagger }E_{\mu
,\nu }=\mathbb{I}$ for all $t$. The operators $E_{\left\{ \mu \right\} }$
describe the one-qubit quantum channel effects.

In the cases where the quantum channel induces a disentanglement only in
asymptotic time, the quantum discord does not disappear in a finite time,
since the entanglement is itself a kind of quantum correlation. Therefore,
we present below what happens to the discord in the ESD situations, for some
of the common channels for qubits: dephasing, generalized amplitude damping
(thermal bath at arbitrary temperature) and depolarizing.

\textit{Dephasing:} The dephasing channel induces a loss of quantum
coherence without any energy exchange \cite{nielsen}. The quantum state
populations remain unchanged throughout the time. To examine the two-qubit
entanglement and discord dynamics under the action of a dephasing channel,
we utilize the Werner state as the initial condition, i.e, $\rho
(0)=(1-\alpha )\mathbb{I}/4+\alpha \left\vert \Psi ^{-}\right\rangle
\left\langle \Psi ^{-}\right\vert $, $\alpha \in \left[ 0,1\right] $ and $%
\left\vert \Psi ^{-}\right\rangle =\left( \left\vert 01\right\rangle
-\left\vert 10\right\rangle \right) /\sqrt{2}$. In this case, we can
calculate analytically the quantum discord in a situation where the
entanglement suddenly disappears. Thus, according to Eq. (\ref{rho}), with
the non-zero Kraus operators for a dephasing channel given by\ $E_{0}=\mbox{%
diag}\left( 1,\sqrt{1-\gamma }\right) $ and \thinspace $E_{1}=\mbox{diag}%
\left( 1,\sqrt{\gamma }\right) $, where $\gamma =1-e^{-\Gamma t}$, $\Gamma $
denoting the decay rate \cite{nielsen}, the elements of the density matrix
of this system evolve to 
\begin{eqnarray*}
\rho _{ii}(t) &=&\rho _{ii}(0),i=1..4, \\
\rho _{23}(t) &=&\rho _{23}(0)\left( 1-\gamma \right) =\rho _{32}(t).
\end{eqnarray*}
The concurrence for this state is given by $C(\rho )=\alpha \left(
3/2-2\gamma \right) -1/2$, which reaches zero in a finite time for any $%
\alpha \neq 1$, as shown in Fig. 1(a). On the other hand, based on the
results given in \cite{luo}, the quantum discord for this state reads $%
D\left( \rho \right) =\left[ F(a+b)+F(a-b)\right] /4-F(a)/2$, where $%
F(x)=x\log _{2}{x}$, $a=(1-\alpha )$ and $b=2\alpha (1-\gamma )$. As shown
in Fig. 1(b), for any $\alpha $, the quantum discord vanishes ($D\left( \rho
\right) =0$) only in the asymptotic limit.
\begin{figure}[!htbp]
\begin{center}
\includegraphics[width=.48\textwidth]{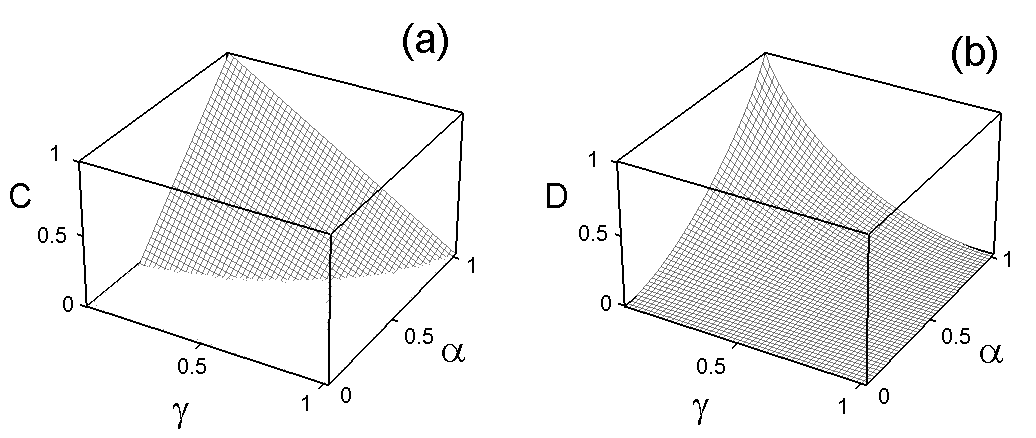} {}
\end{center}
\caption{Dissipative dynamics of (a) concurrence and (b) discord as
functions of $\alpha $ and $\gamma $, assuming independent dephasing
perturbative channels.}
\end{figure}

\textit{Generalized Amplitude Damping (GAD):} The GAD describes the exchange
of energy between the system and the environment, including finite
temperature aspects. It is described by the Kraus operators $E_{0}=\sqrt{q}%
\mbox{diag}\left( 1,\sqrt{1-\gamma }\right) $, $E_{1}=\sqrt{q\gamma }\left(
\sigma _{1}+i\sigma _{2}\right) /2$, $E_{2}=\sqrt{\left( 1-q\right) \gamma }%
\mbox{diag}\left( \sqrt{1-\gamma },1\right) $, and $E_{3}=\sqrt{\left(
1-q\right) \gamma }\left( \sigma _{1}-i\sigma _{2}\right) /2$, where $\gamma 
$ is defined above and $q$ defines the final probability distribution of the
qubit when $t\rightarrow \infty $ ($q=1$ corresponds to the usual amplitude
damping with $T=0K$) \cite{nielsen}.

For the initial condition given by $\rho \left( 0\right) =\left\vert \Phi
\right\rangle \left\langle \Phi \right\vert $ with 
\begin{equation}
\left\vert \Phi \right\rangle =\sqrt{1-\alpha }\left\vert 00\right\rangle +%
\sqrt{\alpha }\left\vert 11\right\rangle ,\text{ \ \ }\alpha \in \lbrack
0,1],  \label{condini}
\end{equation}%
we obtain, according to Eq.(\ref{rho}), the density matrix dynamics 
\begin{subequations}
\label{rho-gad}
\begin{eqnarray*}
\rho _{11}\left( t\right) &=&\rho _{11}\left( 0\right) \left\{ 1-\gamma 
\left[ 2\left( 1-q\right) -\gamma \left( 1-2q\right) \right] \right\}
+\gamma ^{2}q^{2}, \\
\rho _{22}\left( t\right) &=&\rho _{33}\left( t\right) =\gamma \left[ \rho
_{11}\left( 0\right) \left( 1-2q\right) \left( 1-\gamma \right) +q\left(
1-\gamma q\right) \right] , \\
\rho _{44}\left( t\right) &=&1-\rho _{11}\left( t\right) -2\rho _{22}\left(
t\right) , \\
\rho _{14}\left( t\right) &=&\rho _{41}\left( t\right) =\rho _{14}\left(
0\right) \left( 1-\gamma \right) .
\end{eqnarray*}

We examine the dissipative dynamics derived from this channel, taking $q=1$
and $q=2/3$. For these cases, we compute the discord numerically and compare
it with the concurrence. To calculate the discord, we chose the set of
projectors $\left\{ \left\vert \psi _{1}\right\rangle \left\langle \psi
_{1}\right\vert ,\left\vert \psi _{2}\right\rangle \left\langle \psi
_{2}\right\vert \right\} $, where $\left\vert \psi _{1}\right\rangle =\cos {%
\theta }\left\vert 0\right\rangle +e^{i\phi }\sin {\theta }\left\vert
1\right\rangle $ and $\left\vert \psi _{2}\right\rangle =-\cos {\theta }%
\left\vert 1\right\rangle +e^{-i\phi }\sin {\theta }\left\vert
0\right\rangle $, to measure one of the subsystems. The maximum of equation (%
\ref{cc}) is obtained numerically by varying the angles $\theta $ and $\phi $
from $0$ to $2\pi $.

For $q=1$, we have $C(t)=\mathrm{max}\{0,\Lambda _{1}(t)\}$, with $\Lambda
_{1}(t)>0$ for all $t$, whenever $\sqrt{1-\alpha }>\sqrt{\alpha },$see Fig.
2(a). On the other hand, for $q\neq 1$, we have $\Lambda _{1}(t\rightarrow
\infty )<0$ for all $\alpha $, as shown in Fig. 2(c). In this case, since
the concurrence decays monotonically under the Markovian approximation \cite%
{dajka08}, the ESD occurs for any initial superposition state, i.e, for any $%
\alpha $ different from $0$ or $1$.

The discord evidently behaves differently from the concurrence, see Fig.
2(b) and 2(d). In both situations ($q=1$ and $q=2/3$), the discord decays
exponentially and vanishes only asymptotically. For $\alpha =1$ or $\alpha
=0 $ (pure separable states) the quantum discord is zero all the time, as
expected.
\begin{figure}[!htbp]
\begin{center}
\includegraphics[width=.48\textwidth]{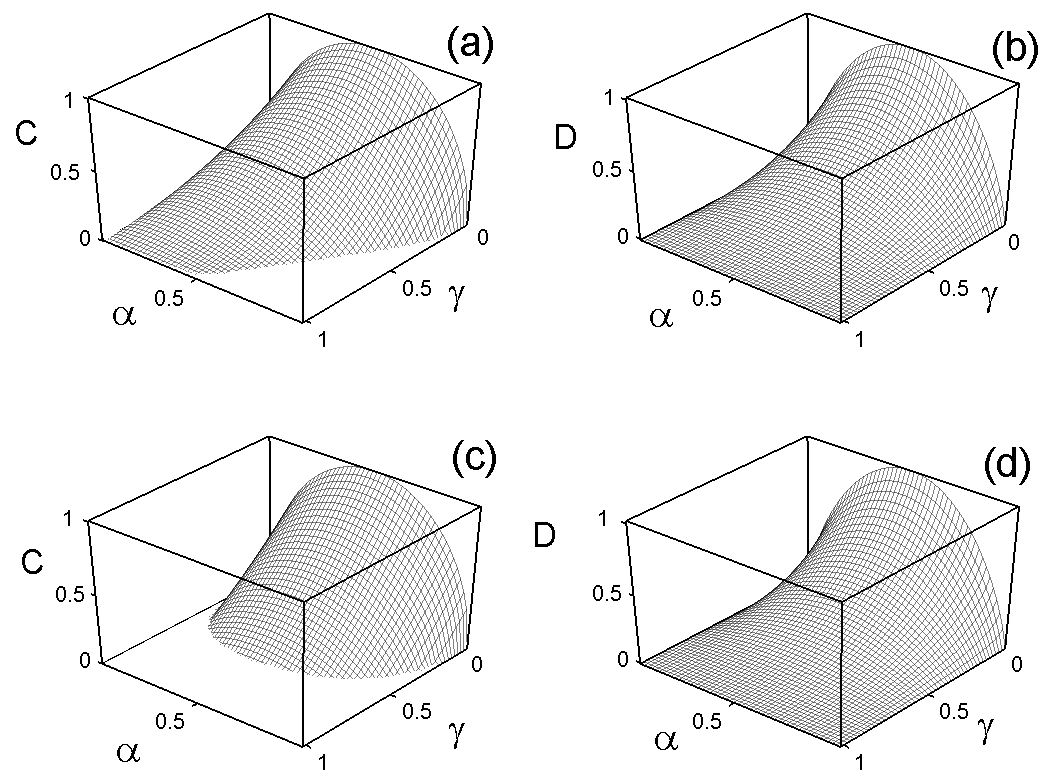} {}
\end{center}
\caption{Dissipative dynamics of concurrence (a,c) and discord (b,d)
as a functions of $\alpha $ and $\gamma $, assuming independent generalized
amplitude damping. (a,b) $q=1$ and (c,d) $q=2/3.$}
\end{figure}

\textit{Depolarizing:} The depolarizing channel represents the process in
which the density matrix is dynamically replaced by the maximal mixed state $%
{\mathbb{I}}/2$, ${\mathbb{I}}$ being the identity matrix of a single qubit.
The set of Kraus operators that reproduces the effect of the depolarizing
channel is given by $E_{0}=\sqrt{1-3\gamma /4}{\mathbb{I}}$, $E_{1}=\sqrt{%
\gamma /4}{\mathbb{\sigma }}_{x}$, $E_{2}=\sqrt{\gamma /4}{\mathbb{\sigma }}%
_{y}$, and $E_{3}=\sqrt{\gamma /4}{\mathbb{\sigma }}_{z},$ with $\gamma $ as
defined above \cite{nielsen}. Assuming the initial condition given by Eq. (%
\ref{condini}), we obtain the density matrix element dynamics: 
\end{subequations}
\begin{eqnarray}
\rho _{11}(t) &=&\rho _{11}(0)\left( 1-\gamma \right) +\gamma ^{2}/4,  \notag
\\
\rho _{22}(t) &=&\rho _{33}(t)=\gamma /2\left( 1-\gamma /2\right) ,  \notag
\\
\rho _{44}(t) &=&1-\rho _{11}(t)-2\rho _{22}(t),  \notag \\
\rho _{14}(t) &=&\rho _{41}(t)=\rho _{14}(0)\left( 1-\gamma \right) ,  \notag
\end{eqnarray}
As in generalized amplitude damping, the ESD occurs for any initial
condition, since $C(t)=\mathrm{max}\{0,\Lambda _{1}(t)\}$ and $\Lambda
_{1}(t\rightarrow \infty )<0$, Fig. 3(a). Again, as shown in Fig. 3(b), the
quantum discord does not disappear in a finite time. Here we used the same
procedure as above to calculate the discord numerically.
\begin{figure}[!htbp]
\begin{center}
\includegraphics[width=.48\textwidth]{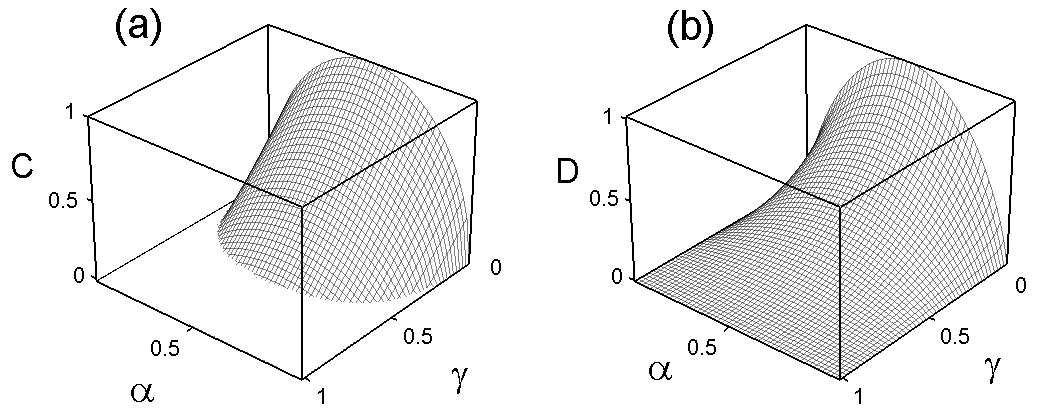} {}
\end{center}
\caption{Dissipative dynamics of (a) concurrence and (b) discord as
functions of $\alpha $ and $\gamma $, assuming independent depolarizing
channels.}
\end{figure}

\textit{Dephasing plus Amplitude Damping:} In Fig 4 we plot the concurrence
and quantum discord for the case where both qubits interact individually
with two distinct reservoirs: those which induce dephasing and amplitude
damping. We assume equal decay rates ($\Gamma $) for both channels, $q=1$ ($%
T=0K$) and $\gamma =1-e^{-\Gamma t}$. For the initial condition given by Eq.
(\ref{condini}), the dephasing channel alone is not able to induce sudden
death of entanglement. On the other hand, in the presence of an amplitude
damping channel, the entanglement suddenly disapears for some values of $%
\alpha $, Fig. 2(a), as discussed above. However, when both channels are
present, the dynamics of the entanglement is very different, suddenly
disappearing for all values of $\alpha $ as we can see in Fig 4(a). This
non-additivity of the decoherence channels in the entanglement dynamics was
firstly pointed out by T. Yu and J. H. Eberly \cite{eberly}, in contrast to
the additivity of de decay rates of different decoherence channels of a
single system \cite{eberly}. But, as shown in Fig 4(b), the discord still
decays asymptotically when both decoherence channels are present, as shown
in Fig 4(b), indicating that the additivity of the decoherence channels is
valid for the quantum discord.
\begin{figure}[!htbp]
\begin{center}
\includegraphics[width=.48\textwidth]{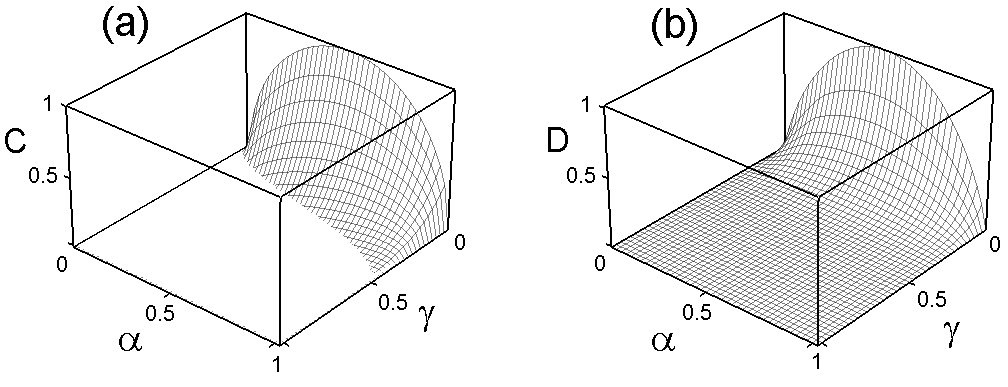} {}
\end{center}
\caption{Dissipative dynamics of (a) concurrence and (b) discord as
functions of $\alpha $ and $\gamma $, assuming simultaneous action of
dephasing and generalized amplitude damping channels in each qubit with $q=1$.}
\end{figure}

In conclusion, we have calculated the discord dynamics for two qubits
coupled to independent Markovian environments. We observed that under the
dissipative dynamics considered here, discord is more robust than
entanglement, even at a finite temperature, being immune to a\ "sudden
death". This also points to a fact\ that the absence of entanglement does
not necessarily indicate the absence of quantum correlations. Thus, quantum
discord might be a better measure of the quantum resources available to
quantum information and computation processes. This also suggests that
quantum computers based on this kind of quantum correlation, differently
from those based on entanglement, are more resistant to external
perturbations and, therefore, introduce a new hope of implementing an
efficient quantum computer. Moreover, discord may be considered, in this
scenario, as a good indicator of classicality \cite{zurek,zurek1}, since it
vanishes only in the asymptotic limit, when the coherence of the individual
qubits disappears. However, we have not demonstrated that the sudden death
of discord in a Markovian regime is impossible, and a study of the discord
from a geometrical point of view \cite{terra-cunha}, for example, might be
useful to address this important question.

We acknowledge helpful discussions with Lucas C\'{e}leri and Marcelo O.
Terra Cunha, and financial support from the Brazilian \ funding agencies
CNPq, CAPES, FAPESP, and the Brazilian National Institute of Science and
Technology for Quantum Information (CNPq).

\end{document}